\documentclass[aip,reprint]{revtex4-1}

\usepackage{lipsum}
\usepackage{amsmath}
\usepackage{amssymb}
\usepackage{graphicx}
\usepackage{gensymb}
\usepackage{tabularx}
\usepackage{float}
\usepackage{xcolor}
\usepackage{mathtools}
\usepackage{cancel}
\usepackage[capitalise]{cleveref}
\DeclareUnicodeCharacter{2009}{ }
\DeclareUnicodeCharacter{2192}{$\rightarrow$}
\DeclareUnicodeCharacter{2032}{'}
\DeclareUnicodeCharacter{03B1}{$\alpha$}
\DeclareUnicodeCharacter{03B3}{$\gamma$}
\restylefloat{table}

\begin{document}
	\title{SpK: A fast atomic and microphysics code for the high-energy-density regime}
	\author{A. J. Crilly}\email{ac116@ic.ac.uk}
	\affiliation{Centre for Inertial Fusion Studies, The Blackett Laboratory, Imperial College, London SW7 2AZ, United Kingdom}
	\author{N. P. L. Niasse}
	\affiliation{Centre for Inertial Fusion Studies, The Blackett Laboratory, Imperial College, London SW7 2AZ, United Kingdom}
	\affiliation{First Light Fusion Ltd., Unit 9/10 Oxford Industrial Park, Mead Road, Yarnton, Kidlington OX5 1QU, United Kingdom}
	\author{A. R. Fraser}
	\affiliation{Centre for Inertial Fusion Studies, The Blackett Laboratory, Imperial College, London SW7 2AZ, United Kingdom}
	\affiliation{First Light Fusion Ltd., Unit 9/10 Oxford Industrial Park, Mead Road, Yarnton, Kidlington OX5 1QU, United Kingdom}
	\author{D. A. Chapman}
	\affiliation{First Light Fusion Ltd., Unit 9/10 Oxford Industrial Park, Mead Road, Yarnton, Kidlington OX5 1QU, United Kingdom}
	\author{K. W. McLean}
	\affiliation{Centre for Inertial Fusion Studies, The Blackett Laboratory, Imperial College, London SW7 2AZ, United Kingdom}
	\author{S. J. Rose}
	\affiliation{Centre for Inertial Fusion Studies, The Blackett Laboratory, Imperial College, London SW7 2AZ, United Kingdom}
	\author{J. P. Chittenden}
	\affiliation{Centre for Inertial Fusion Studies, The Blackett Laboratory, Imperial College, London SW7 2AZ, United Kingdom}
	
	\begin{abstract}
		SpK is part of the numerical codebase at Imperial College London used to model high energy density physics (HEDP) experiments. SpK is an efficient atomic and microphysics code used to perform detailed configuration accounting calculations of electronic and ionic stage populations, opacities and emissivities for use in post-processing and radiation hydrodynamics simulations. This is done using screened hydrogenic atomic data supplemented by the NIST energy level database. An extended Saha model solves for chemical equilibrium with extensions for non-ideal physics, such as ionisation potential depression, and non thermal equilibrium corrections. A tree-heap (treap) data structure is used to store spectral data, such as opacity, which is dynamic thus allowing easy insertion of points around spectral lines without a-priori knowledge of the ion stage populations. Results from SpK are compared to other codes and descriptions of radiation transport solutions which use SpK data are given. The treap data structure and SpK's computational efficiency allows inline post-processing of 3D hydrodynamics simulations with a dynamically evolving spectrum stored in a treap.
	\end{abstract}
	
	\maketitle
	
	\section{Introduction}\label{section:intro}
	
	Radiation magnetohydrodynamics (RMHD) codes are commonly used in the modelling of High Energy Density Physics (HEDP) experiments. The system of RMHD equations require external data/models to be closed, namely opacities, equation of state and transport coefficients. Constructing synthetic diagnostics from RMHD simulations requires post-processing with accurate models of the emission, absorption and scattering processes. Microphysics models are used to calculate thermodynamic, transport and radiative properties and their accuracy is vital in HEDP modelling. 
	
	In microphysics models, a thermodynamic ensemble of quantum mechanical states must be considered in order to determine macroscopic properties. Marrying a small scale accurate quantum mechanical description with the many-body and many-species thermodynamic system is an exceedingly difficult task and a field of active research. There are two overarching approaches; the `physical' and `chemical' pictures. In the physical picture, one works at the level of electrons and nuclei, calculating electronic states self-consistently. Bound electronic states cause ions, atoms and molecules to emerge. However, current physical picture models cannot independently describe all of parameter space due to the approximations made in their construction or due the computational expense or intractability. The approach that we shall use is in the chemical picture. In the chemical picture, we define species within the material; free electrons, bound electrons, ions, molecules, etc. Each of these species is described through its own free energy and interaction term which can be derived from the partition functions.
	
	A benefit of the chemical picture is that it is conceptually closer to the macroscopic quantities we wish to extract. Chemical picture models can be constructed to work over large parts of density-temperature parameters space and are generally computationally expedient. However, there are two issues which arise from this approach. Firstly, although the equations of chemical equilibrium are simpler than the many-body equations of the physical picture, one must construct models for the partition functions based on microscopic properties and interactions. For all but the simplest terms, these models are approximate in nature. Secondly, we have drawn a distinction between chemical species such as free and bound electrons where physically there is no difference in their properties and interactions. Indeed, core bound electrons and high energy free electrons do behave differently and are well described by different models. However, in the middle ground it is less clear that the distinction is meaningful. Interactions between free and bound electrons must be added back into the model after the separation of terms -- a physical picture model has no such issue.
	
	In constructing a suitable microphysics model, we wish to be able to compute the quantities needed to close the RMHD equations across a wide range of parameter space in multi-material systems and have the potential to run inline to post-process RMHD simulations to produce synthetic diagnostics. Therefore, a balance between fidelity, efficiency and robustness must be struck in our choice of models. For example, accounting for the fine structure of multi-electron ions in a plasma background is a great computational task so a reduced atomic model must be used to allow efficient calculations. As discussed above, a chemical picture model is best suited to the task of describing a wide parameter space and in plasma physics the Saha equation\cite{Saha1920} is the canonical example of this. In this work, we describe the fast atomic and microphysics model SpK, which has been developed at Imperial College London to be used in conjunction with the RMHD code Chimera\cite{Chittenden2016,McGlinchey2018,Tong2019}. SpK is an atomic model code which uses the screened hydrogenic model (SHM) with energy levels taken from the NIST database to perform detailed configuration accounting (DCA) calculations of the state populations, free electron number density, emissivity and opacity at given conditions using an extended Saha model. SpK can work in local thermodynamic equilibrium (LTE) or collisional-radiative equilibrium (CRE), the latter allowing the modelling of low density, mid Z plasmas. Multigroup emissivity and opacity tables produced by SpK are used within the radiation-hydrodynamics code Chimera and SpK inline or tables are used post-process simulations to produce synthetic diagnostics. 
	
	\section{Atomic Physics Model}
	\subsection{Energy levels and configurations}\label{section:atomicmodel}
	
	As a balance between fidelity and efficiency, SpK uses detailed configuration accounting (DCA) with levels split by quantum numbers $n$ and $l$. This makes for an efficient structuring where a given bound electron configuration of the form:
	\begin{equation*}
	(nl_1^{N_{l_1}})(nl_2^{N_{l_2}})(...)(n'l_1'^{N_{l_1'}})(n'l_2'^{N_{l_2'}})(...)
	\end{equation*}
	constitutes a singular energy level. The energy of the configuration can either be calculated using a simplified model or looked up from the NIST energy level database. In SpK, the default is to use NIST energy level values but when these are unavailable the screened hydrogenic model with $l$-splitting developed by Faussurier \textit{et al.}\cite{Faussurier2008} is used. In this model, each subshell electron experiences a $Z_k/r$ central potential where $Z_k$ is a constant screened nuclear charge. This screened charge is calculated using the occupation of the other subshells and pre-computed screening coefficients ($\sigma_{kk'}$):
	\begin{equation}
	Z_k = Z-\sum_{k'} \sigma_{kk'}(N_{k'}-\delta_{kk'})
	\end{equation}
	where $N_{k}$ is the occupancy of the $k$-th subshell and $\delta$ is the Kronecker delta to prevent self-screening. Single electron wavefunctions have a hydrogenic form but scaled by the effective charge. This allows analytic formulae to be derived for important quantities such as dipole matrix elements. In summary, the implemented DCA atomic model uses the NIST database, when available, for energy levels and screened hydrogenic model results for other atomic properties, such as dipole matrix elements and oscillator strengths. In this way, SpK's atomic model is suitably complete without requiring any explicit numerical solution to the many-body Hamiltonian.
	
	The approach described above provides an efficient atomic model from which macroscopic quantities such as opacity can be calculated. However, this has come at the sacrifice of accuracy. The lack of atomic fine structure has a large impact on the level of spectroscopic detail. For example, as levels are split only on $n$ and $l$, the $1s 2p \rightarrow 1s^2$ transition is not split into resonance and inter-combination lines. The use of the screened hydrogenic model to calculate energy levels and matrix elements can lead to errors over more sophisticated Hartree-Fock calculations or measurements. Additionally, relativistic effects\cite{Cowan1981} are appreciable for atomic charges $>$ 30 and, while NIST energy levels will include these effects, screened-hydrogenic-model predictions are largely classical and thus increasingly erroneous at larger $Z$. Therefore, the SpK atomic model's current range of validity for nuclear charges is $Z < 30$. Relativistic extensions to the screened hydrogenic model\cite{Mendoza2011} could be used to extend this range.
	
	\subsection{Radiative processes}\label{section:radiative}
	
	In section \cref{section:populationsolver} we will describe the population solver used to calculate the ionic and electronic state populations. In this section, we will assume knowledge of the electronic structure and populations and show how the radiative properties of plasmas can be calculated using the atomic physics models described in \cref{section:atomicmodel}. As we are working in the chemical picture, free and bound electrons are treated differently. Therefore, we will consider free-free, bound-free and bound-bound electronic transitions separately. We will also assume thermal equilibrium in the following.
	
	The free-free process, bremsstrahlung, occurs within the free continuum of states so can produce/absorb photons at any energy. The opacity of thermal bremsstrahlung for ion species $i$ is given by \cite{Longair2011}:
	\begin{align}
	\kappa^{\mbox{ff}}_{i,\nu} &= n_i \sigma_{i,\mbox{ff}} \left[1-\exp\left(-\frac{h\nu}{k_BT}\right)\right] g_{\mbox{ff}}(\nu,T) \phi_{\mbox{ff}}(\nu,T,\eta) \ ,
	\end{align}
	where $\sigma_{i,\mbox{ff}}$ is the semi-classical cross section from Kramers\cite{Kramers1923}, $g_{\mbox{ff}}$ is the free-free Gaunt factor and $\phi_{\mbox{ff}}$ is the free-free degeneracy correction\cite{Rose1993}. An electron transition between a bound and free state can produce photons at any energy above a threshold. The threshold energy is the energy required to promote the electron from the bound state to the lowest energy free state. The opacity of thermal bound-free absorption for state $j$ is given by \cite{Rybicki2008}:
	\begin{align}
	\kappa^{\mbox{bf}}_{j,\nu} &= n_j \sigma_{j,\mbox{bf}}  \left[1-\exp\left(-\frac{h\nu}{k_BT}\right)\right] g_{j,\mbox{bf}}(\nu,T)\phi_{j,\mbox{bf}}(\nu,T,\eta) \ ,
	\end{align}
	where $\sigma_{j,\mbox{bf}}$ is the semi-classical photo-ionisation cross section for state $j$, $g_{j,\mbox{bf}}$ is the bound-free Gaunt factor and $\phi_{j,\mbox{bf}}$ is the bound-free degeneracy correction\cite{Rose1993}. In SpK, photo-ionisation cross sections are calculated using the semi-classical formula from Rose\cite{Rose1998}. Gaunt factors are introduced to include full quantum-mechanical corrections to the semi-classical cross section expressions. In SpK, Karzas \& Latter\cite{Karzas1961} free-free and bound-free Gaunt factors are calculated using the code from Janicki\cite{Janicki1990}.
	
	The opacity for the line transition between states $j$ and $k$ is: 
	\begin{align}
	\kappa^{\mbox{bb}}_{jk,\nu} &= \frac{h\nu_{jk}}{4\pi} n_j B_{jk} \left[1 - \frac{g_jn_k}{g_kn_j}\right] \label{eqnbbopacity_nonLTE} \ ,\\
	\frac{g_jn_k}{g_kn_j} &= \exp\left[-\frac{h\nu_{jk}}{k_BT}\right] \ ,
	\end{align}
	it should be noted that \cref{eqnbbopacity_nonLTE} is true even out of thermodynamic equilibrium. The Einstein coefficients can be translated into the more familiar oscillator strength, $f_{jk}$, notation. The oscillator strength quantifies the quantum-mechanical correction to the classical expression based on the Larmor formula:
	\begin{align}
	B_{jk} &= \frac{\pi e^2}{\epsilon_0m_ech\nu_{jk}}f_{jk} \Phi(\nu) \ ,\\
	f_{jk} &= \frac{2}{3}\frac{m_e}{\hbar^2}h\nu_{jk} \left|r_{jk}\right|^2 \ ,
	\end{align}
	where $r_{jk}$ is the dipole matrix element and $\Phi(\nu)$ is the line shape function. In SpK, dipole matrix elements are calculated using the screened hydrogenic expressions of Khandelwal \textit{et al.}\cite{Khandelwal1989}.
	
	In SpK, the line shape function, $\Phi(\nu)$, takes a Voigt profile due to the convolution of Gaussian and Lorentzian shapes from various physical processes. The Gaussian width is taken as the sum of Doppler broadening and the unresolved transition array (UTA) width. The Lorentzian width is taken as the sum of the natural line width and Stark broadening. A fast and accurate numerical expression from Limandri \textit{et al.}\cite{Limandri2008} is used to compute the Voigt shape function.
	
	Fast approximate models for Stark broadening, such as Dimitrijevic and Konjevic\cite{Dimitrijevic1987,Dimitrijevic1980}, use the Born-Bethe inelastic cross section based on dipole matrix elements. The total electron impact Lorentzian broadening for a transition between states $i$ and $k$ is given by\cite{Griem1968}:
	\begin{align}
	\gamma^{EI}_{ik} &= k_{EI} \left[\sum_j g_{EI}(\nu_{ij})\left|r_{ij}\right|^2 + \sum_l g_{EI}(\nu_{kl})\left|r_{kl}\right|^2\right]\ , \\
	k_{EI} &= 3\left(\frac{4 \pi E_H}{3}\right)^{\frac{3}{2}} \frac{n_e a_0^3}{\sqrt{k_BT_e}} \ ,
	\end{align}
	where $g_{EI}$ are the Maxwellian averaged electron impact Gaunt factors as given by Griem\cite{Griem1968} and Dimitrijevic and Konjevic\cite{Dimitrijevic1987,Dimitrijevic1980}. Again, screened hydrogenic dipole matrix elements are used in this calculation.
	
	Within multi-electron systems, a configuration to configuration transition is potentially split into a large number of term lines depending on the arrangement of component angular momenta. DCA calculations are not resolved at the term level so miss these additional line structures, this leads to large inaccuracies in opacity evaluations. One successful method to compensate for these missing term lines is using unresolved transition arrays (UTAs). In UTA models, statistical moments of the term levels are used to smooth the single DCA configuration to configuration transition to better represent the true structure of term lines. In SpK, the distribution energy moments of Bauche, Arnoult and Peyrusse (Table 3.2 in reference\cite{Bauche2015}) are used to calculate the configuration energy widths. These calculations involve Slater integrals, for which we use Naqvi's screened hydrogenic formulae\cite{Naqvi1963}, Wigner 3- and 6-j symbols, for which we use the SLATEC subroutines\cite{Abramowitz1972,Messiah2014,Schulten1975a,Schulten1975b,Schulten1976}, and finally the spin-orbit integrals, for which we used the screened hydrogenic analytic result\cite{Cowan1981}. The UTA Gaussian width is calculated as the quadrature sum of upper and lower level configuration widths\cite{Rose1992,Moszkowski1962}:
	\begin{equation}
	\sigma_{\mathrm{UTA}}^2 = \sigma_{E,i}^2+\sigma_{E,j}^2 \ , \label{eqn:UTAwidth}
	\end{equation}
	where $i$ and $j$ are the upper and lower level indices and $\sigma_E$ is the configuration width. It is noted that for a more detailed UTA model\cite{Bauche2015} the width of the transition array is generally narrower than the quadrature sum of configuration widths. Summing the configuration widths equates to the lowest order UTA model as it assumes no correlation between the levels and ignores selection rules from LS coupling. The correlation arises from the propensity law that high (low) energy levels of the upper configuration de-excite preferentially towards the high (low) levels of the lower configuration to, approximately, respect core and spin invariance selection rules\cite{Bauche2015}. Future work will look to improve the SpK UTA model. 
	
	Natural line broadening arises due to the uncertainty principle and contributes to a Lorentzian line shape. Canonically, the line width is given as the sum of spontaneous emission Einstein coefficients. On top of this natural width, one can add the radiation broadening due to stimulated emission. Including both photo-excitation and photo-ionisation, the full width formula for a transition from upper state $i$ is given by\cite{Mclean2021}:
	\begin{align}
	\gamma_{RB} &= \frac{2\pi h e^2}{\epsilon_0 m_e c^3} \left[\sum_{j<i}\frac{2l_j+1}{2l_i+1}\nu_{ij}^2\left(1+\frac{c^3}{8\pi h \nu_{ij}^3}E(\nu_{ij})f_{ij}\right) \right. \nonumber \\
	&+ \sum_{k>i}\frac{c^3}{8\pi h \nu_{ik}}E(\nu_{ik})f_{ik} \nonumber \\
	&\left.+ \int_I^\infty \frac{hc^3}{8\pi} \frac{df_i}{d(h\nu)} E(\nu) dh\nu \right] \ ,
	\end{align}   
	where $E(\nu)$ is the frequency resolved radiation energy density, $I$ is photo-ionisation energy of state $i$ and $df_i/d(h\nu)$ is the differential oscillator strength. In SpK, the differential oscillator strength is assumed $\propto \nu^3$ and the constant of proportionality is calculated using the f-sum rule\cite{Bethe2012}. Radiation broadening is only a significant effect in low electron density plasmas in high radiation temperature backgrounds\cite{Mclean2021}, it is therefore not routinely included in SpK calculations.
	
	\section{Spectral Data Structures}
	
	Multi-electron and/or multi-material systems involve many energy levels leading to large numbers of transitions. High energy resolution is desired around spectral lines but not required in the continuum regions. When calculating a spectrum, it is not known a-priori which ionic stages and electronic states will be occupied and therefore which transitions will be present in the spectrum. However, it can be very memory-expensive to include resolution around every possible transition. Thus when calculating spectra, the question of data storage is crucial. The most common way to record the properties of an object set involves the use of a data structure called an array. A typical array is defined as a contiguous zone of the computer memory that stores a collection of indexed elements. When a value needs to be saved or retrieved from a location, the address of the corresponding memory zone is calculated and directly accessed in $\mathcal{O}$(1) time. If the number of objects to insert is not known before runtime, a growable or dynamic array is needed. Such a structure is created by estimating at the very beginning the maximum amount of memory to be required and by allocating an array of such fixed-size. Elements can then be added or removed at the end of this reserved space in constant time. As simple as this approach is, it has several disadvantages. Firstly, if the initial estimate of the maximum size $n$ is smaller than the number of elements to be finally inserted, a reallocation of the array (taking $\mathcal{O}$($n$) time) has to be performed when full capacity is reached. Secondly, any insertion or deletion at an arbitrary position in the array takes a linear time as it requires the following elements to be moved. Finally, if no simple formula links the index of the record with its value, searching for an arbitrary element may necessitate looping through every single object in the array. 
	
\begin{figure}[htp]
		\centering
		\includegraphics*[width=0.95\columnwidth]{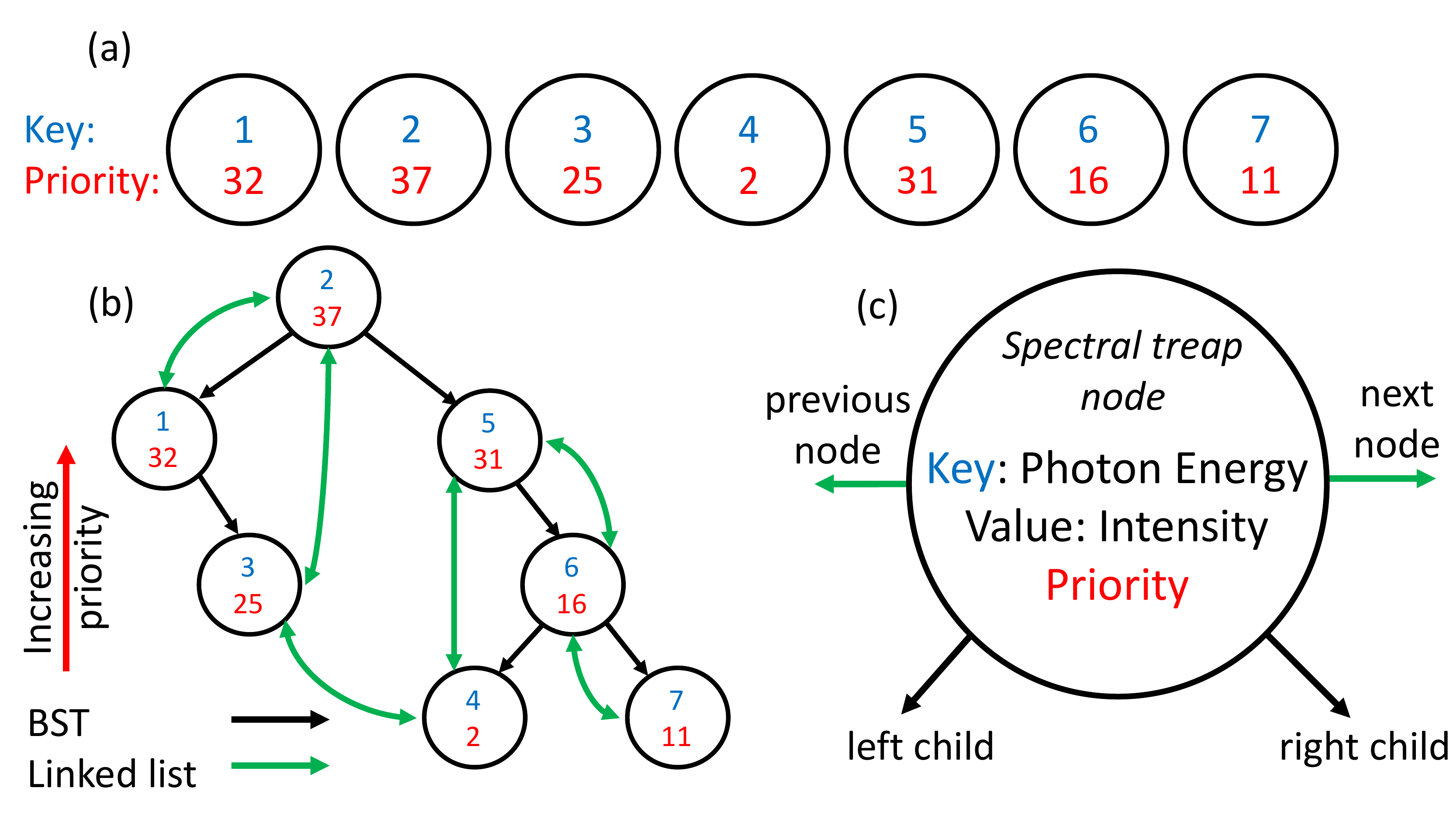}
		\caption{(a) A set of unstructured nodes with key values ($\in$ [1,7]) and random priority values. The priorities set an insertion order of 2, 1, 5, 3, 6, 7 and 4. (b) The constructed treap with both a key-ordered binary search tree and heap-ordered priorities. Also shown is an additional doubly linked list which links adjacent key values. (c) A schematic of the spectral treap node structure implemented in SpK. }
		\label{fig:TreapDiagram}
	\end{figure}

	To address the issues with static arrays for spectral data storage, a method was derived from a self-balancing binary search tree implementation. A tree is a data structure using an organization of nodes mimicking an arborescent hierarchy. Each node is a parent object that may have several children located directly below it in the structure. The tree begins at a single node known as the `root' and nodes with no children are known as `leaves'. Binary trees are a special case of tree topology in which each node has a maximum of two children, one `left' and one `right'. Along with the local node data, the memory addresses of those children are stored in the pointer field of the node object. A Binary Search Tree (BST) is a binary tree with special features. The data zone of a BST node contains a field for a so-called key value. This key belongs to an ordered set with a comparison operator. Searching the tree involves comparing to the key value and moving left or right depending on the result of the comparison, and repeating until the matching key is found. A BST is said to be balanced if none of its leaves is much farther away from the root than the others. For such a balanced BST, the average number of steps between the root and one leaf is log$_2 (n)$. Search and insert procedures can be very quick on balanced BSTs. Unfortunately, in general, there is no guarantee that a generic construction process will lead to such a balance. The treap data structure was first introduced by Aragon and Seidel\cite{Aragon1989} in 1989. Although similar in many respects to BSTs, treaps have the particular ability to remain balanced. Compared to regular tree nodes, treap node objects have an extra field containing a randomly generated value called priority. This value, allocated at the node creation, determines its place in the hierarchy: any operation (insertion/deletion) performed on the treap, while keeping the in-order traversal order (key order) intact, must ensure that the parents have higher priorities than their children. The final structure obtained by following this rule is equivalent to a binary tree constructed by inserting the items in decreasing order of priority (i.e. priority decreasing with distance from the root). If the initial distribution of node priorities is random, the tree is balanced with high probability. The averaged $\mathcal{O}$(log $n$) time complexity for search, insertion and deletion operations makes treaps ideal candidates for spectral data storage. In our approach, each spectral point is represented by a node whose key and value are respectively equal to the photon frequency and spectral intensity. Operations performed on line profiles often required working on a set of points with good frequency locality. A broadening process, for example, will modify the amplitude of the spectral points on the left and right of the line center. If we start by pointing to the node at this central frequency, devising a method to provide fast access to the nearby nodes is desirable. A more practical solution consists in maintaining a doubly linked list on top of the treap by adding two extra pointers to the node object which point to the nodes with adjacent key values. An example of a treap structure and its construction is given in \cref{fig:TreapDiagram}.
	
	The dynamic nature of the treap data structure is particularly well suited to the solution of radiative transfer in 3D hydrodynamic simulations. The spatial variation in density and temperature will give rise to varying emission and absorption spectra. Being able to dynamically add resolution around emission and absorption features as one propagates a spectrum through the simulation grid gives both accuracy and efficiency. SpK's post-processing capabilities will be discussed further in \cref{section:postprocessing}.
	
	\section{Population Solver}\label{section:populationsolver}
	
	In order to evaluate macroscopic properties such as opacity, the occupation of free and bound electron states must be calculated. We will assume that the atomic kinetic time-scale is much faster than the evolution of the macroscopic system, and thus an equilibrium will be reached. The solution of the equilibrium electron population will depend on thermodynamic properties such as density and temperature. In the following we will describe the theory and numerical models used in SpK for this calculation.
	
	\subsection{Local Thermal Equilibrium, LTE}
	
	In thermal equilibrium, partitioning of a thermodynamic system between the different species is determined through chemical equilibrium. For plasmas, we are concerned with the equilibrium between free electrons and the various ion stages. For the ideal plasma, this is found with the Saha equation, where one considers the chemical equilibrium of ionisation between stages $i$ and $i+1$:
	\begin{align}
	A_i &\rightleftharpoons A_{i+1}+e^- \ , \\
	\mu_i &= \mu_{i+1}+\mu_e \ , \label{eqnchemicaleq}
	\end{align}
	where $\mu$ is the chemical potential and the subscript denotes the species. Using the canonical expressions for the ideal translational and internal partition functions, the expression of chemical equilibrium becomes:
	\begin{align}
	\frac{N_{i+1} N_e}{N_{i}} &= N_e e^{-\beta \mu_e}  \left(\frac{\mathcal{Q}_{int,i+1}}{\mathcal{Q}_{int,i}}\right) \\
	&= N_e e^{-\beta \mu_e} \left(\frac{\sum_n g_{n,i+1} \exp \left[-\beta \epsilon_{n,i+1} \right]}{\sum_n g_{n,i} \exp \left[-\beta \epsilon_{n,i} \right]}\right) e^{-\beta I_i} \label{eqnSahaEquation} \ , \nonumber
	\end{align}
	where $\beta = 1/k_BT$ is the inverse temperature, $I_i$ is the ground state ionisation energy, and $g_{n,i}$ and $\epsilon_{n,i}$ are the state degeneracies and excitation energies of state $n$ for ion stage $i$. The omitted ionic translational partition functions cancel almost exactly as the ion masses only differ by a single electron mass. The chemical potential of the free electrons can be calculated for a given $N_e$, in the classical limit this simplifies to:
	\begin{equation}
	\lim_{\mu_e\to-\infty}\left(\frac{N_{i+1} N_e}{N_{i}}\right) = \frac{2V}{\lambda_e^3} \left(\frac{\mathcal{Q}_{int,i+1}}{\mathcal{Q}_{int,i}}\right) \ ,
	\end{equation}
	where $V$ is the volume and $\lambda_e$ is the electron thermal deBroglie wavelength. We see increased temperature drives ionisation and increased density drives recombination. While the Saha solution is well-behaved over the whole space, one must remember that here we have neglected all interactions between species, such as the Coulomb attraction between free electrons and ions. Therefore, the present Saha result is inaccurate at higher densities. Even when neglecting interaction terms, one will quickly run into difficulties. This is because the internal partition function diverges if the infinity of bound levels of an isolated atom/ion are considered. This non-physicality is a symptom of the separate treatment of bound and free electrons. In reality, the highest lying bound states will be perturbed by nearby charges and become delocalised. The truncation of the internal partition function is an active field of research including the phenomena of ionisation potential depression (IPD) and electric microfields.
	
	\subsubsection{Non-ideal effects}
	
	The combined effects of pressure ionisation can be included in our chemical picture model by modifying the internal partition function. Following the work of Mihalas, Hummer and D\"appen \cite{MHD1,MHD2,MHD3}, this is done through the reduced statistical weights, $w_n$, and shift to the continuum, $\Delta$. Mathematically these effects are given by:
	\begin{align}
	\mathcal{Q}_{int} &\rightarrow \sum_n w_ng_n \exp \left[-\beta E_n \right] \ ,\\
	I_i &\rightarrow I_i - \Delta \ .
	\end{align}
	This assumes the bound states continue to resemble the corresponding isolated atom states and that the dominant effects are an effective lowering of the continuum and reduction of bound state occupation near the continuum. This is a good approximation for core electrons but can expected to be poorer for levels nearer the continuum. More sophisticated models, such as INFERNO\cite{Liberman1979} and PURGATORIO\cite{Wilson2006}, solve self-consistently the `ion in jellium' system and thus resolve the individual level response.
	
	Various IPD models exist and each uses different approximations to arrive at the shift in the continuum. A popular model developed by Stewart and Pyatt \cite{SP1966} retrieves the low coupling ($\Gamma = \frac{1}{3}\kappa_D^2r_{ws}^2 \ll 1$, where $r_{ws} = (3/4\pi n_i)^{1/3}$) Debye limit and high coupling ion-sphere limit. The Stewart-Pyatt IPD model has had some success \cite{Hoarty2013} and some failure \cite{Ciricosta2012} in explaining experimental data. In SpK, one can switch between the Debye, ion-sphere or Stewart-Pyatt analytic models, although Stewart-Pyatt is used by default. 
	
	Recent work by Lin \textit{et al.}\cite{Lin2017,Lin2019}, has shown that the static structure factor (SSF or $S(k)$) can be used to calculate IPD. These quantum statistical models are attractive as they offer a more rigorous and accurate calculation of IPD. Also, established non-ideal statistical mechanics models such as the Hyper-Netted-Chain (HNC) approximation can be used to calculate the SSF. Lin \textit{et al.}\cite{Lin2017,Lin2019} showed one can split the IPD contributions from fast moving free electrons and the ions surrounded by their screening cloud of slow moving electrons. The calculation of the ionic contribution to the continuum shift is reduced to an integral over the screened ionic SSF\cite{Lin2017,Lin2019}:
	\begin{align}
	\Delta^{\mathrm{ion}} &= \frac{e^2}{2 \pi^2 \epsilon_0 r_i^2} F(\Gamma_i) \int_{0}^{\infty} \frac{S_{ii}^{ZZ}(q)}{q^2} dq \ , \\
	S_{ii}^{ZZ}(k) &= \left(1-\frac{q_{sc}(k)}{\bar{Z}}\right)^2 S_{ii}(k) \ . \label{eqn:SiiZZ}
	\end{align}
	Similarly, for the free electron contribution:
	\begin{align}
	\Delta^{\mathrm{elec}} &= \frac{e^2}{2 \pi^2 \epsilon_0 r_i^2} F(\Gamma_i) \int_{0}^{\infty} \frac{S_{ee}^0(q)}{q^2} dq \ ,
	\end{align}
	where we define the following mean charges for the reduced system:
	\begin{align}
	\bar{Z} = \frac{\sum Z_i n_i}{\sum n_i} \ \ , \ \ Z^* = \frac{\sum Z_i^2 n_i}{\sum Z_i n_i} \ ,
	\end{align}
	and coupling strength $\Gamma_i$, ionic radius $r_i$ and the $F(\Gamma_i)$ function are defined as\cite{Lin2019}:
	\begin{align}
	\Gamma_i &=  \frac{(\bar{Z}+1)(Z^*+1)e^2}{4\pi \epsilon_0 r_i k_B T} , \\
	r_i &= \left[\frac{3(\bar{Z}+1)}{4\pi n_e}\right]^{\frac{1}{3}} \ , \\
	F(\Gamma_i) &= \frac{3 \Gamma_i}{\sqrt{1-0.4\left(3\Gamma_i\gamma_0^2\right)^{3/4}+3\Gamma_i\gamma_0^2}} \ , \gamma_0 \equiv \left[\frac{4}{9\pi}\right]^{\frac{1}{3}} \ .
	\end{align}
	The IPD for a given ion stage $i$ is then given by the charge-scaled sum of these components:
	\begin{equation}
	\Delta_i = \frac{Z_i+1}{Z^*+1}\left[Z^*\Delta^{\mathrm{ion}}+\Delta^{\mathrm{elec}}\right]
	\end{equation}
	Towards including a similar model in SpK, a one-component-plasma (OCP) HNC code has been implemented to numerically calculate the screened ionic SSF. The Ng acceleration scheme\cite{Ng1974} and the separation of long- and short-ranged terms given by Springer \textit{et al.}\cite{Springer1973} are utilised. The OCP HNC equations are solved for the electron-screened\cite{Kremp2006} interaction between average-ions of charge $\bar{Z}$ in order to obtain the SSF. Separate to the HNC code, the electron screening cloud function, $q_{sc}(k)$, is calculated including the static local field correction, $G_{ee}$:
	\begin{align}
	q_{sc}(k) &= \frac{-\bar{Z}\Pi^{R}_{ee}(k)V_{ee}(k)}{1 - (1-G_{ee})\Pi^{R}_{ee}(k)V_{ee}(k)} \ ,
	\end{align}
	where $V_{ee}(k) = e^2/(\epsilon_0k^2)$ is Fourier-space electron-electron Coulomb potential, $\Pi^{R}_{ee}$ is the retarded polarisation function and is calculated using the closed form expressions from Dandrea \textit{et al.}\cite{Dandrea1986}, and $G_{ee}$ is calculated using the Farid \textit{et al.}\cite{Farid1993} expression. Currently, the free electron SSF, $S_{ee}^0(q)$, is taken as the Debye limiting case. While this IPD model has improved accuracy over Stewart-Pyatt and similar models, it is less robust and considerably more computationally expensive. Results from this model will be discussed in detail in \cref{section:IPD}.
	
	The pressure ionisation (PI) model used to calculate the reduced statistical weights, $w_n$, is based on the work of Mihalas, Hummer and D\"appen \cite{MHD1,MHD2,MHD3,Nayfonov1999}. Using the microfields distribution, one calculates the probability that a state remains bound in the presence of the microfields. The critical electric field which unbinds a state is the product of the classical saddlepoint field\cite{MHD3} and a Stark ionisation factor\cite{MHD3}, $k_n$. The microfields distribution, $P(\beta)$, is used to find the cumulative probability, $w_{n}^{m}$, of experiencing a field strength less than a critical field, $F^c$:
	\begin{align}
	w_{n}^{m} &= \int_0^{\beta_n^c} P(\beta) d\beta, \ \ \beta_n^c = k_n \frac{F_n^c}{F_0} \ ,\\
	F_n^c &= \frac{\pi \epsilon_0 E_n^2}{Ze^3}, \ \ F_0 = \frac{\bar{Z}e}{4\pi\epsilon_0 r_{ws}^2} = \frac{en_e}{4\pi\epsilon_0 n_i}\left(\frac{4 \pi n_i}{3}\right)^{2/3} \ ,
	\end{align}
	where $F_n^c$ is the saddle point field for state $n$ of energy $E_n$; and $F_0$ is the field strength arising from an ion at the mean interionic separation. SpK has implementations of both the Holtsmark\cite{Poquerusse2000} and Q-fit\cite{Nayfonov1999} microfield distributions. At high densities, bound states can also be perturbed strongly by frequent overlap with other extended particles. Mihalas, Hummer and D\"appen also reduce the statistical weight of states the more volume they occupy relative to that remaining after taking into account the volumes occupied by all other atoms and ions. To save considering all possible excited states of each ion in the calculation of volume available to a single state, the approximation is made that all perturbing ions are in their ground state\cite{MHD1,MHD2,MHD3,Nayfonov1999}. SpK also includes this model, which gives the probability a state exists as:
	\begin{equation}
	w_{n}^{g} = \exp\left(-\frac{4\pi}{3}\sum_{i}n_{i}(r_{i}^{g}+r_{n})^{3}\right).
	\end{equation}
	Here, $r_{n}$ and $r_{i}^{g}$ are the expectation values of the radii of the outermost orbitals of state $n$ and the ground state of ion species $i$, $n_{i}$ is the total number density of the ion species, and the summation takes place over all possible ion species (including neutral atoms). The overall occupation probability of a state is then the product of probabilities arising due to both plasma microfields and ground-state perturbers, $w_{n}=w_{n}^{m}w_{n}^{g}$.
	
	\subsubsection{Numerical scheme}
	
	The logarithm of the Saha equations are solved iteratively using a Picard step with damping \cite{Rouse1961}, the logarithm equations more accurately capture the disparate scales which occur in the ionic stage populations. With ionic stage index $i$ and iteration index $j$, the Saha iterative loop has the following form \cite{Niasse2012}:
	\begin{align}
	n_{e,j} &= \bar{Z}_j n_{tot} \ ,\\
	\theta_{i+1,j} &= \theta_{i,j} + \ln\left[\mathcal{Q}_{e,j}\right] + \ln \left[\frac{\mathcal{Q}_{int,i+1,j}}{n_{e,j}\mathcal{Q}_{int,i,j}}\right]-\beta(I_i-\Delta_{i,j}) \label{eqnSpKSaha} \ ,\\
	f_{i,j}       &= \exp\left[\theta_{i,j} - \mbox{max}_i(\theta_{i,j})\right] \ ,\\
	\bar{Z}'_j &= \frac{\sum_i Z_i f_{i,j}}{\sum_i f_{i,j}}, \ \ n_{i,j} = \frac{f_{i,j}}{\sum_i f_{i,j}} n_{tot} \ ,\\
	\bar{Z}_{j+1} &= \lambda \bar{Z}_{j}+(1-\lambda) \bar{Z}'_j \label{eqnSpKPicard} \ ,
	\end{align}
	where $n_{tot}$ is the total number of nuclei in all ionic stages, $\mathcal{Q}$ denotes the internal and free electron partition functions and \cref{eqnSpKSaha} is the logarithm of the Saha equations. The ionic `fractions', $f_{i,j}$, are constructed such that they have a maximal value of 1. This is to ensure stability over a large range of ion densities. Picard iteration (\cref{eqnSpKPicard}) updates the $j$th ionisation estimate where the damping, $\lambda$, is typically set at 90\%. To initialise the iterative Saha solution, the average ionisation is estimated using a Thomas-Fermi estimate \cite{More1985,Rose1992}, i.e. $\bar{Z}_{0} = \bar{Z}_{TF}$. The Saha loop is exited when $\bar{Z}_{j+1}$ and $\bar{Z}_{j}$ converge within tolerance.
	
	\subsection{Non Local Thermal Equilibrium, NLTE}
	
	Busquet \textit{et al.}\cite{Busquet1982,Busquet1993,Busquet2006,Busquet2009} have developed a numerically simple method to transform an existing LTE code into a pseudo non-LTE (NLTE) one, where simple modifications to the Saha equation can be used to capture NLTE effects. To illustrate this methodology, let us first consider a simple atomic configuration consisting of only two single levels of respective population densities $n_1$, $n_2$ statistical weights $g_1$, $g_2$ and energy separation $\Delta E$. In equilibrium, the populations and rates $W$ between levels balance:
	\begin{equation}
	\frac{n_2}{n_1} = \frac{W_{1\rightarrow 2}}{W_{2\rightarrow 1}} = \frac{n_e C_{12}+4\pi I_\nu B_{12}}{n_e C_{21}+A_{21}+4\pi I_\nu B_{21}}
	\end{equation}
	with $C_{ij}$ the rate of electron collisional excitation/de-excitation between the two levels $i$ and $j$. The different rates can be obtained using the expression of the Einstein coefficients and following the estimation of the collisional excitation cross section given by Van Regemorter\cite{vanRegemorter1962} and Mewe\cite{Mewe1972}. For collisional-radiative equilibrium, we only consider collisional excitation/de-excitation and spontaneous radiative decay:
	\begin{align}
	\frac{n_2}{n_1} &= \frac{g_2}{g_1}\frac{\exp\left[-\beta \Delta E\right]}{1+\frac{2\sqrt{3}}{c^3h^2e^2\bar{g}}\sqrt{\frac{\pi k_BT_e}{2m_e}}\frac{\Delta E^3}{n_e}} \ ,
	\end{align}
	where $\bar{g}$ is the Gaunt factor for collisional excitation. The above expression tends towards the Boltzmann distribution for high electron density, low electron temperature and small energy gap. In other cases, the upper level is depopulated in comparison to the Boltzmann equilibrium. Busquet \textit{et al.} showed that one can extend this expression in order to evaluate the equilibrium populations between two ion stages $i$ and $i+1$, summing up the transition rates between the superlevels formed by the excited state configurations of each ion and averaging the transition energies:
	\begin{align}
	\frac{n_{i+1}}{n_{i}} &= \frac{\sum W_{i \rightarrow i+1}}{\sum W_{i+1 \rightarrow i}} \ , \\
	&= \frac{2}{n_e\lambda_e^3}\left(\frac{\sum_n g_{n,i+1} \exp \left[-\beta \epsilon_{n,i+1} \right]}{\sum_n g_{n,i} \exp \left[-\beta \epsilon_{n,i} \right]}\right) \\
	&\times \frac{\exp\left[-\beta \langle\Delta E\rangle\right]}{1+\frac{2\sqrt{3}}{c^3h^2e^2\bar{g}}\sqrt{\frac{\pi k_BT_e}{2m_e}}\frac{1}{n_e}\frac{\langle\Delta E^2 f\rangle}{\langle\Delta E^{-1}f \bar{g}\rangle}} \ , \nonumber
	\end{align}
	where $f$ is the oscillator strength and angle brackets denote occupation probability averages over the states of the ion stage.
	
	Finally, using the following approximation was found to not significantly change the resulting ionisation equilibrium:
	\begin{align}
	\frac{\langle\Delta E^2 f\rangle}{\langle\Delta E^{-1}f \bar{g}\rangle} \approx 5 \cdot I_{i}^3
	\end{align}
	where $I_i$ is the ionisation energy of ion stage $\alpha$. Combining these results gives a simple multiplicative factor to the Saha equation:
	\begin{align}
	\frac{n_{i+1} n_e}{n_{i}} &= f_{\mathrm{Saha,LTE}}(n_e,T_e) K_{\mathrm{NLTE}}(n_e,T_e) \ , \\
	K_{\mathrm{NLTE}} &= \left[1+\alpha \frac{I_i^3 \sqrt{k_BT_e}}{n_e}\right]^{-1} \ ,
	\end{align}
	where $\alpha = 1.34 \times 10^{13}$ cm$^{-3}$ eV$^{-7/2}$. This additional multiplicative term is equivalent to including a modified electron temperature in exponent of the LTE Saha equation:
	\begin{align}
	e^{-\frac{I_i}{k_BT_e}}K_{\mathrm{NLTE}} &= e^{-\frac{I_i}{k_BT_{e,\mathrm{eff}}}} \ , \\
	T_{e,\mathrm{eff}} &= \frac{T_e}{1+\frac{T_e}{I_i}\ln\left(1+\alpha \frac{I_i^3 T_e^{1/2}}{n_e}\right)} \ ,
	\end{align}
	showing that the NLTE effects act as an effective reduction in the electron temperature, with the greatest effect for ion stages with large ionisation energies e.g. H-like and He-like ions. This formulation is also used to modify expressions for emissivity and opacity by considering the altered level populations in NLTE.
	
	\section{Model Comparisons}
	
	In the following subsections, we will compare SpK results against analytic and computational models from the literature. While SpK operates in SI units (with eV for temperatures), we will facilitate easy comparisons by converting to the units used in the literature.
	
	\subsection{Ionic Fractions}
	
\begin{figure}[htp]
		\centering
		\includegraphics*[width=0.95\columnwidth]{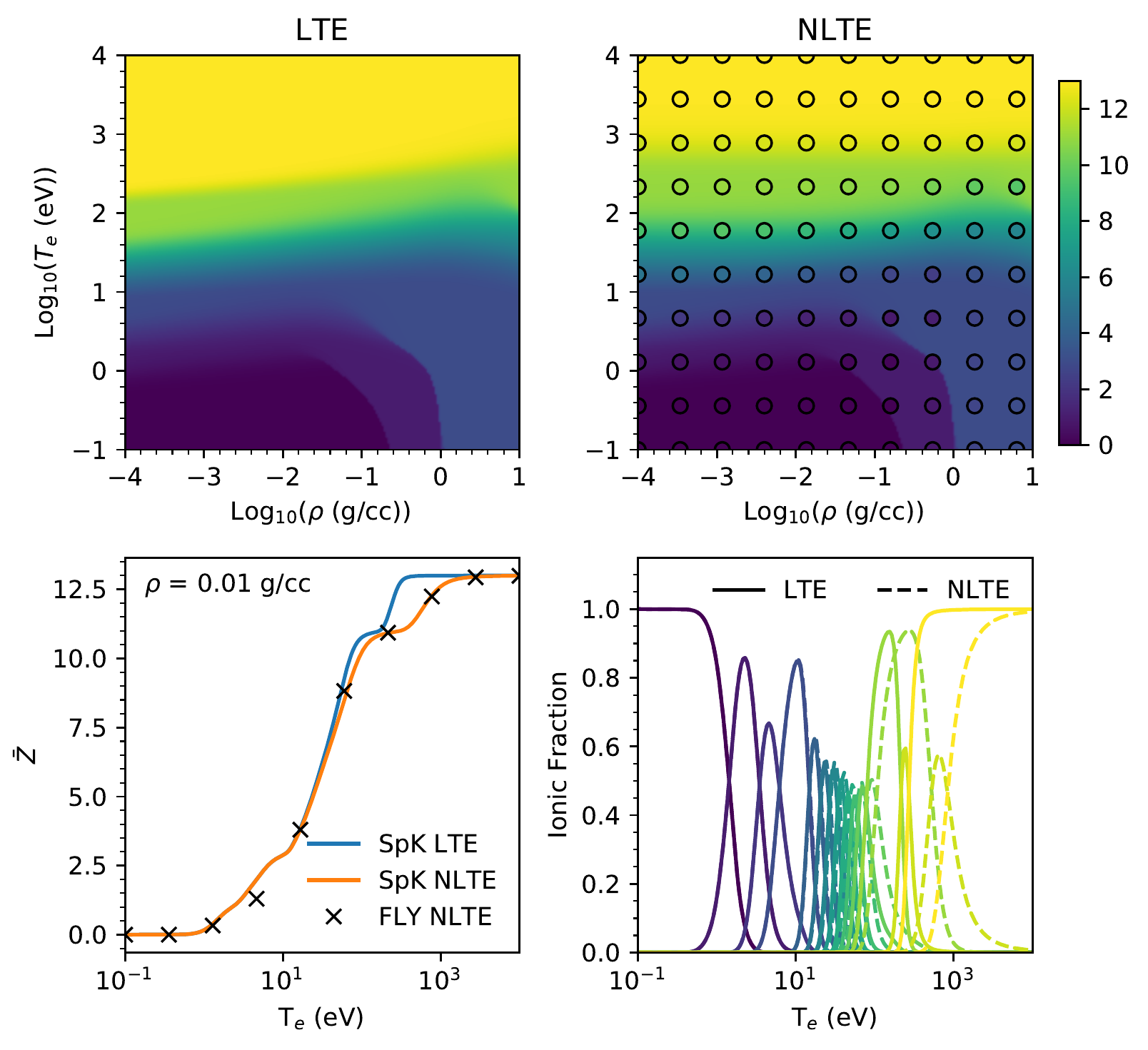}
		\caption{(Top left) LTE SpK calculation of $\bar{Z}$ for Al. (Top right) NLTE SpK calculation of $\bar{Z}$ for Al. Overlaid as coloured circles are results from a steady state NLTE FLYCHK calculation. (Bottom left) Temperature lineout of $\bar{Z}$ at mass density of 0.01 g/cc. (Bottom right) Fractional occupation of each ionic stage as calculated by SpK in LTE and NLTE for a mass density of 0.01 g/cc. The largest deviations between LTE and NLTE are seen at high temperature and high ionic stages.}
		\label{fig:Zbar_SpK_FLY}
	\end{figure}

	The average ionisation, $\bar{Z}$, and the ionic fractions are fundamental in the calculation of macroscopic plasma properties such as the equation of state, transport coefficients and opacity. Common average atom models, such as Thomas-Fermi, do not resolve the separate ionic stages. NLTE effects can also have a significant impact on the ionisation equilibrium and the resulting derived quantities. In this section we will compare SpK's calculation of ionisation to the NLTE code FLYCHK\cite{FLYCHK}. FLYCHK utilises the screened hydrogenic model and allows solution to the full rate equations in its NLTE calculations. In this comparison, the steady state optically thin NLTE solution from FLYCHK is used. We will use Aluminium as a test case - the results are shown in \cref{fig:Zbar_SpK_FLY}.
	
	Good agreement is seen between SpK and FLYCHK on the average ionisation level. The largest deviations are seen at intermediate densities ($\lesssim \rho_{\mathrm{solid}}$) and low temperatures ($\sim 1$ eV). This warm dense matter (WDM) regime is highly sensitive to the pressure ionisation models so disagreement is to be expected.
	
	\subsection{Integrated Opacities}
	
\begin{figure}[htp]
		\centering
		\includegraphics*[width=0.95\columnwidth]{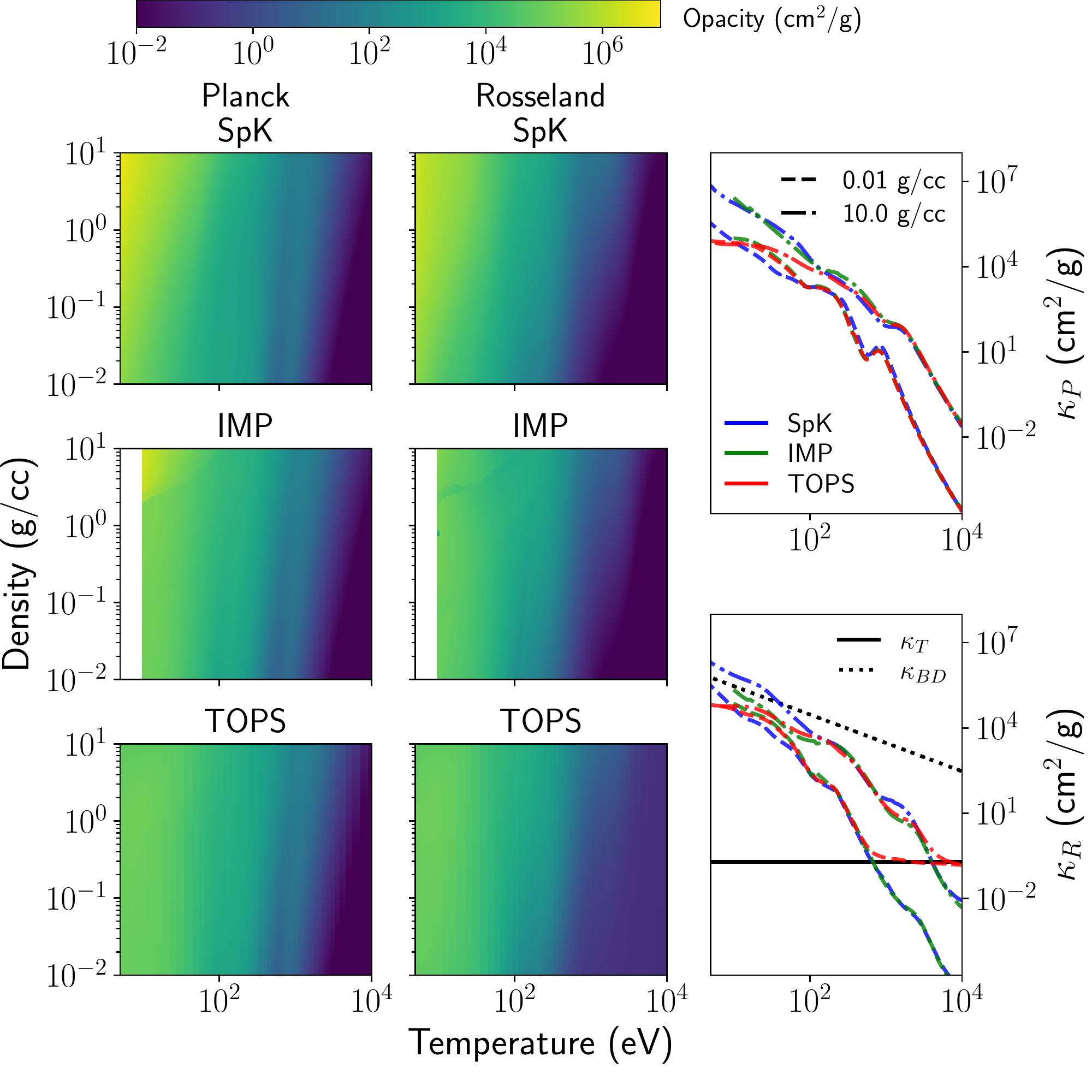}
		\caption{(Left column) Planck opacities of Fe$_2$O$_3$ as calculated by SpK, IMP and TOPS. IMP calculations failed below a temperature of 10 eV. (Middle column) Rosseland opacities of Fe$_2$O$_3$ as calculated by SpK, IMP and TOPS. (Top Right) Lineouts of the Planck mean opacity at 0.01 and 10.0 g/cc as a function of temperature. (Bottom Right) Lineouts of the Rosseland mean opacity at 0.01 and 10.0 g/cc as a function of temperature. Also shown are, in the solid black lines, the scattering opacity $\kappa_T$ calculated from the Thomson cross section and, in the dotted black line, the Bernstein \& Dyson opacity upper limit \cite{Berstein2003}, $\kappa_{BD}$. The plotted TOPS Rosseland opacities include the scattering opacity, while SpK and IMP do not.}
		\label{fig:Plk_Ros_comparison}
	\end{figure}

	The Planckian and Rosseland mean opacities are key coefficients in the transport of radiation through a medium. Within the diffusive approximation, the Planckian opacity describes the absorption/emission properties of the plasma and the Rosseland opacity sets the diffusivity of the radiation. These mean opacities also serve as a good benchmark as the whole of density-temperature phase space can be easily visualised. Model deviations suggest differences/errors in the population and radiative property calculations. In this section, we will compare opacities for Iron Oxide (Fe$_2$O$_3$) given its use in HEDP radiative properties experiments \cite{Hoarty2019,McLean2022}. LTE Planckian and Rosseland opacities from SpK, TOPS\cite{Magee1995} and IMP\cite{Rose1992} are compared in \cref{fig:Plk_Ros_comparison}.
	
	Good agreement is found across the temperature range between all models at lower densities. The models separate more at higher densities, with SpK predicting the highest integrated opacities at low temperature. This is to be expected as SpK's approach is to modify the ideal Saha solution with additional non-ideal physics terms, which will likely break down at high coupling where we are far from ideal gas behaviour. In fact, SpK breaks the Bernstein \& Dyson Rosseland opacity upper limit at low temperatures. SpK's inaccuracy at high coupling parameter can be remedied by interpolating to cold opacity data - this method is described in \cref{section:coldopacities}.
	
	\subsection{Spectrally Resolved Opacities}
	
	SpK can also produce spectrally resolved opacities which fully utilise the strengths of the dynamic treap data structure. While SpK's atomic model is often insufficient for spectroscopic quality calculations of spectral lines, intermediate spectral resolution is very important for synthetic diagnostics and RMHD modelling. Filtered X-ray diagnostics depend on the incident spectrum and multigroup radiation transport requires multigroup opacity tables, which will be described further in \cref{section:tables}.
	
\begin{figure}[htp]
		\centering
		\includegraphics*[width=0.9\columnwidth]{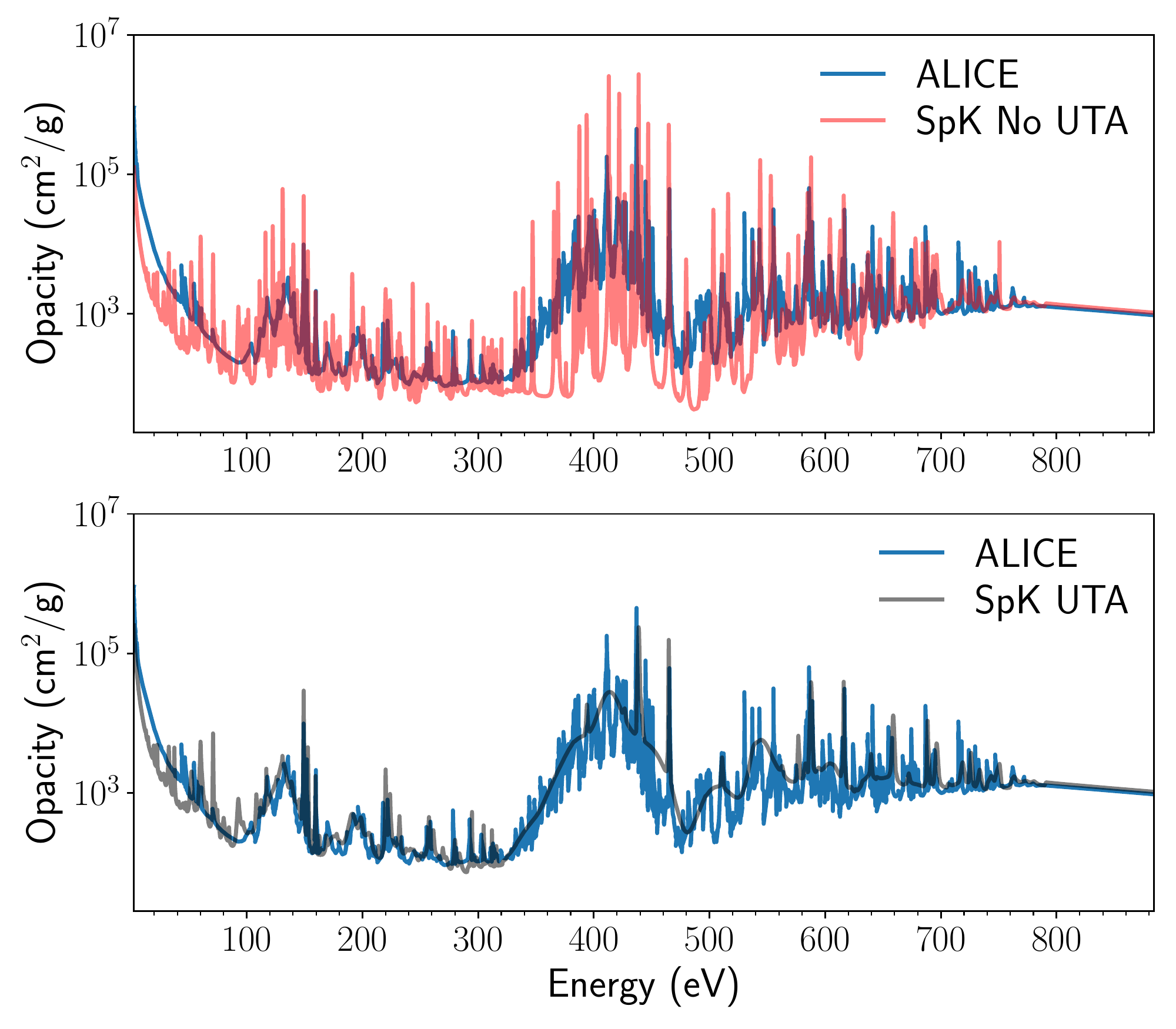}
		\caption{SpK LTE spectrally-resolved opacity calculation for Chlorine at 0.005 g/cc and 100 eV comparing to results from the detailed term accounting code ALICE \cite{Hill2018}. The separate subplots show SpK calculations with and without the UTA model described in \cref{section:radiative}. The inclusion of UTAs ensures good agreement for the average opacity within the line forests. }
		\label{fig:UTA_SpK-Alice}
	\end{figure}

	In this section, we will compare SpK results against the detailed term accounting code ALICE \cite{Hill2018}. We will use the test case used by Hill \textit{et al.}\cite{Hill2018} of an LTE opacity calculation for Chlorine at 0.005 g/cc and 100 eV, for which ALICE was compared to other well-benchmarked opacity codes. \Cref{fig:UTA_SpK-Alice} shows a comparison between SpK and ALICE calculations for this test case.
	
	Without the UTA model SpK has too few spectral lines due to DCA lacking fine structure, this leads to troughs of low opacity between lines. To approximately account for the fine structure line forests, the DCA spectral lines are broadened by an additional Doppler width calculated using UTA theory, see \cref{eqn:UTAwidth}. When the UTA model is used within SpK, good model agreement on the spectrally resolved opacity is obtained. However, SpK does not resolve the line forests and instead provides a much smoother spectrum.

	\subsection{Ionisation Potential Depression}\label{section:IPD}
	
\begin{figure}[htp]
		\centering
		\includegraphics*[width=0.95\columnwidth]{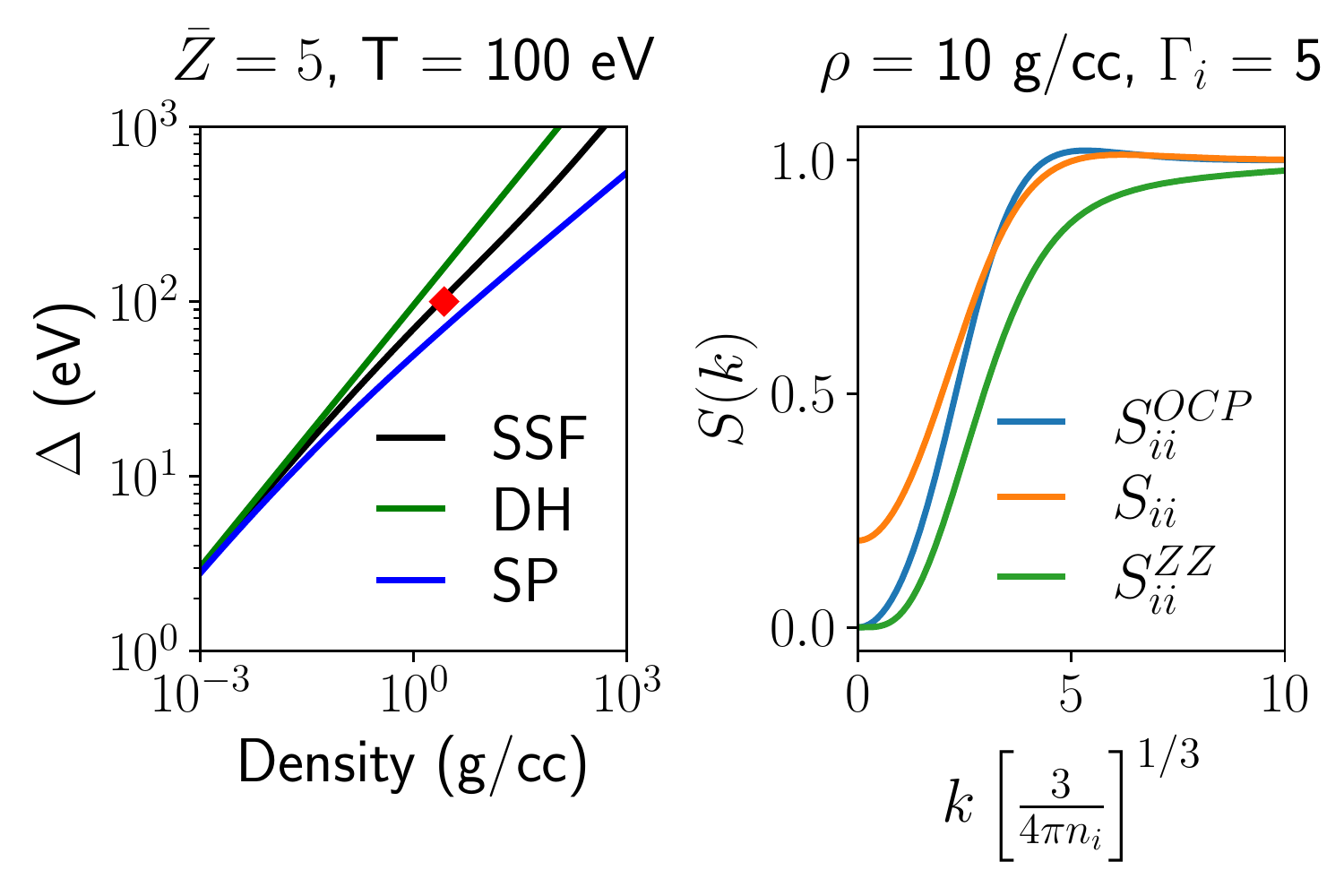}
		\caption{(Left) IPD energies as a function of density for a $\bar{Z}$ = 5 and $T$ = 100 eV. Results from the static structure factor (SSF), Debye-H\"{u}ckel (DH) and Stewart-Pyatt (SP) models are shown. Also shown as a red diamond is the experimental result from Ciricosta \textit{et al.}\cite{Ciricosta2012} for 100 eV Al at 6$\times$10$^{22}$ 1/cc number density. (Right) Corresponding HNC calculations of the SSF at 10 g/cc, $\bar{Z}$ = 5 and $T$ = 100 eV. The $S_{ii}^{OCP}$ is calculated using unscreened ion potentials in the HNC and a screening cloud $q_{sc} = 0$; $S_{ii}$ and $S_{ii}^{ZZ}$ are as defined in \cref{eqn:SiiZZ}.}
		\label{fig:IPD_Sii_Al5_100eV}
	\end{figure}

	As the plasma coupling parameter increases, non-ideal effects such as IPD will be become more significant. Accurate IPD models are particularly important in modelling the warm dense matter regime where the coupling parameter is order unity. SpK has a variety of IPD models implemented, the most sophisticated of which is the static structure factor based calculation derived by Lin \textit{et al.} \cite{Lin2017,Lin2019}. In this section, we will compare IPD models and evaluate their effect on ionisation. In \cref{fig:IPD_Sii_Al5_100eV} we show calculations of the IPD from various models for a single ionic species plasma. The SSF calculation lies between the ideal Debye-H\"{u}ckel and Stewart-Pyatt results across a wide range of densities. This is to be expected as Stewart-Pyatt interpolates between Debye-H\"{u}ckel and the large coupling ion sphere limit. Previous work on SSF IPD has shown the Stewart-Pyatt model tends to the ion sphere limit much faster than more detailed calculations\cite{Lin2017}. The IPD calculated by the SSF model is in line with the experimental result of Ciricosta \textit{et al.}\cite{Ciricosta2012}. Also shown in \cref{fig:IPD_Sii_Al5_100eV} are the SpK HNC calculations of the SSF. Large differences are observed between the unscreened one component plasma $S_{ii}$ and the charge-charge SSF, $S_{ii}^{ZZ}$, showing the electron screening cloud has a non-negligible effect on the low frequency ionic part of the charge fluctuations \cite{Gregori2007}.
	
\begin{figure}[htp]
		\centering
		\includegraphics*[width=0.99\columnwidth]{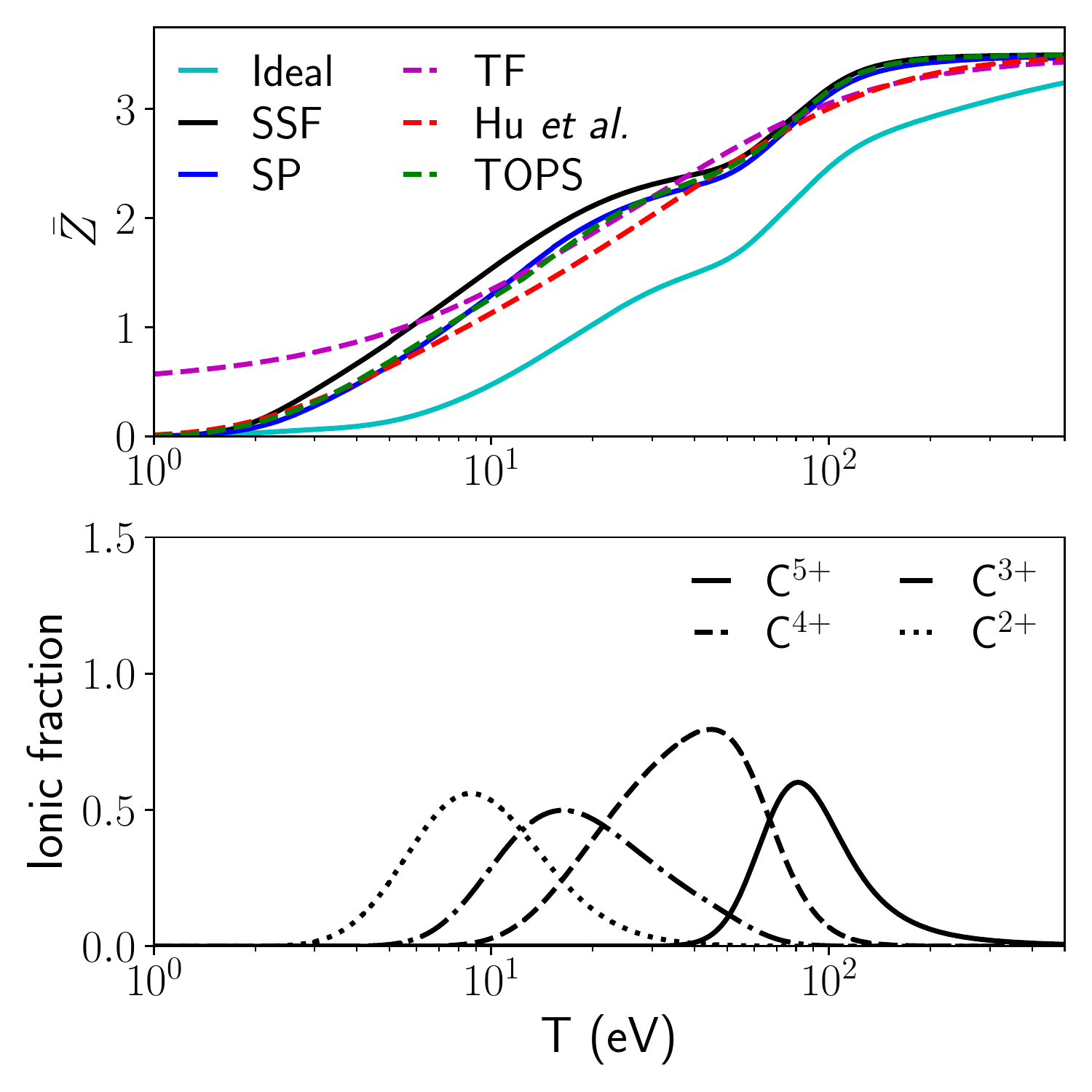}
		\caption{(Top) Average ionisation of CH at 0.1 g/cc as a function of temperature. Lines are predictions from various models: Thomas-Fermi (TF) in dashed magenta, Hu \textit{et al.}'s `Saha-type' fit\cite{Hu2016} to orbital-free molecular dynamics simulations in dashed red, TOPS\cite{Magee1995} in dashed green, the no IPD or `ideal' prediction from SpK in cyan, the SSF-based IPD SpK result in black and the Stewart-Pyatt (SP) result in blue. Both SpK results use the Q-fit microfields pressure ionisation model. (Bottom) The ionic fractions for various charge states of Carbon as a function of temperature from the SSF SpK calculations.}
		\label{fig:IPD_CH}
	\end{figure}

	With SpK, we can also investigate how the different models affect the ionic stage populations by coupling the IPD models to the Saha equilibrium solver. We will investigate the ionisation equilibrium for polystyrene (CH) at 10\% of solid density. Polystyrene is a common ablator material in inertial confinement fusion (ICF) capsules and thus accurate calculation of its properties are key for predictive modelling of ICF experiments. The results of this investigation are shown in \cref{fig:IPD_CH}. It is clear that including pressure ionisation physics is necessary in this regime as the ideal (no IPD) result predicts a significantly lower ionisation than other models. The shell structure is evident in the SpK calculations and leads to a more structured change in ionisation as a function of temperature. Both the SSF and Stewart-Pyatt IPD models were used in conjunction with the Q-fit microfields model of pressure ionisation. Using the Holtsmark distribution leads to much larger deviations as the C L-shell was ionised at much lower temperatures. This is to be expected as screening reduces the probability of experiencing large local electric fields so the ideal gas Holtsmark treatment is inappropriate for WDM. The ionisation of the CH is in good agreement with TOPS, especially for the Stewart-Pyatt model.
	
	While the first results from the SSF-based IPD model are encouraging, future work is needed to ensure it can be used as a standard model in SpK over analytic models like Stewart-Pyatt. The model itself is not sufficiently robust. To ensure convergence in the HNC model, the density must be stepped up and the radial distribution function iteratively evolved to gain stability. At very high coupling, the HNC code can fail to converge completely. The free-electron SSF, $S^0_{ee}$, can also be more accurately evaluated, whether through the random phase approximation (RPA) or including the dynamic local field correction, $G(k,\omega)$. The model could also be extended to also calculate the microfields distribution, allowing a self consistent picture of the non-ideal dense plasma physics. Until this capability is developed, we have shown the combination of the Stewart-Pyatt IPD and Q-fit microfields models are suitably accurate and can be used across the whole parameter space. 
	
	\section{Opacity Tables \& Radiation Transport}\label{section:tables}
	
\begin{figure}[htp]
		\centering
		\includegraphics*[width=0.9\columnwidth]{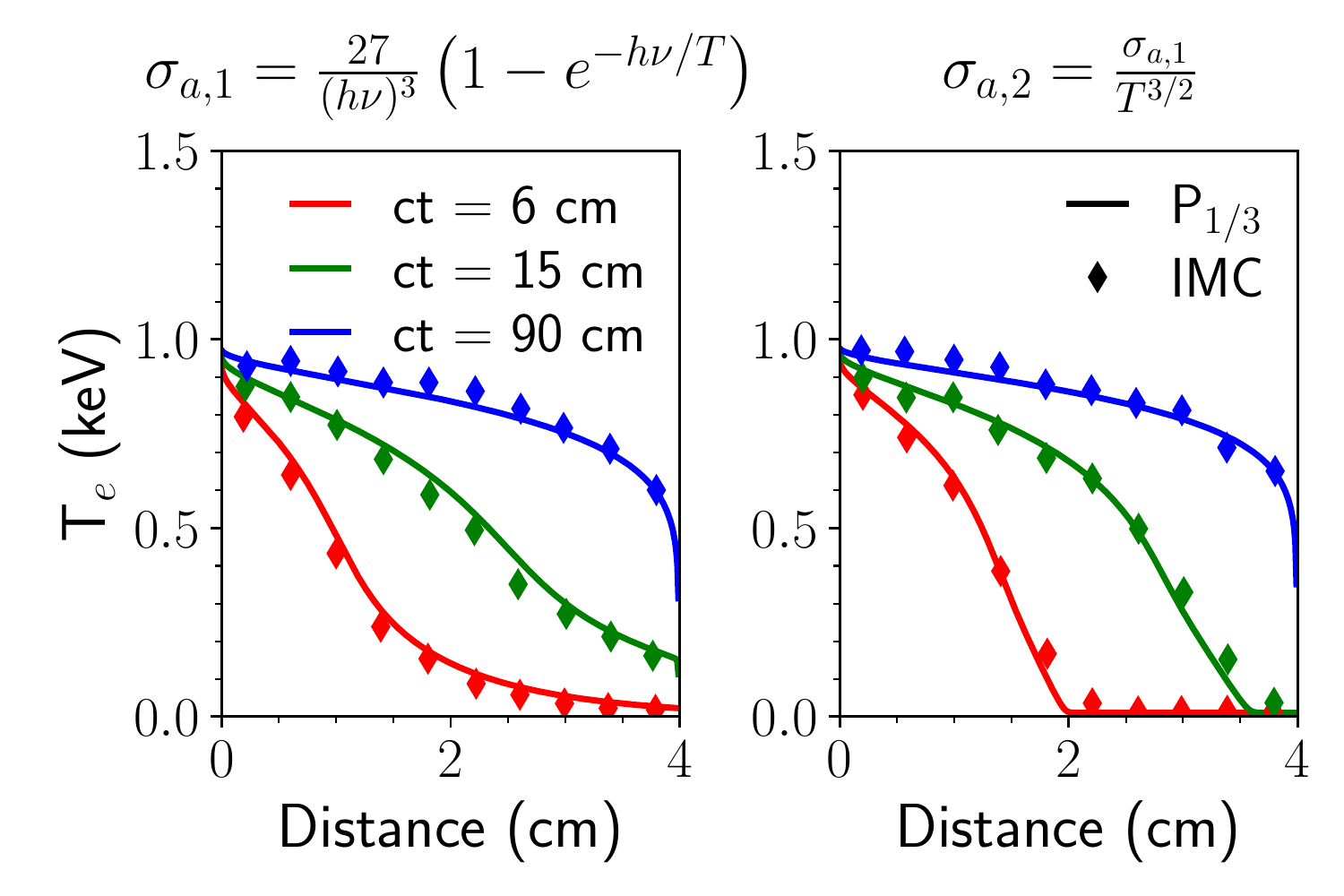}
		\caption{A numerical benchmark of Chimera with Fleck \& Cummings' implicit Monte Carlo solution to frequency dependent radiation transport, compared at 3 different times, $t$, given as distance travelled in the free-streaming limit, $ct$. Chimera is shown in the solid lines and Fleck \& Cummings' results are shown with the diamond symbols. The two panels show the results for two different absorption cross sections (in cm$^{-1}$, with $h\nu$ and $T$ in keV) proposed in the original work\cite{Fleck1971}. The plotted Chimera simulations used a cell resolution of $100$ um and 100 photon energy groups.}
		\label{fig:FleckCummings}
	\end{figure}

	One of the primary uses of SpK is to produce LTE and NLTE opacity tables for use in (RMHD) simulations using the code Chimera\cite{Chittenden2016}. Chimera is an Eulerian radiation-magnetohydrodynamics code with $P_{1/3}$\cite{Olson2000} multigroup radiation transport\cite{Stone1992}, flux-limited Spitzer-Härm thermal transport\cite{Spitzer1953} with explicit super-time-stepping\cite{Meyer2014}, extensive extended MHD capabilities\cite{Walsh2017}, Monte-Carlo alpha particle transport\cite{Tong2019} and equation of state tables from FEoS\cite{FEOS1,FEOS2}. The $P_{1/3}$ multigroup radiation transport equations are as follows:
	\begin{align}
	\frac{\partial E_\nu}{\partial t} &= \kappa_{P,\nu} \left( S_\nu -  c E_\nu\right) - \nabla \cdot \vec{F}_\nu , \\
	\frac{\partial \vec{F}_\nu}{\partial t}  &= - 3 \kappa_{F,\nu} c' \vec{F}_\nu -  c' \nabla (cE_\nu) \ , \\
	\frac{dU_e}{dt} &=  \int  \kappa_{P,\nu} \left( c E_\nu - S_\nu \right)d\nu \ .
	\end{align}
	where we have included a reduced speed of light, $c'$, which can be used to increase computational efficiency at the cost of accuracy in the free-streaming limit and the flux-averaged opacity\cite{Stone1992}, $\kappa_F$. The explicit time integration is performed using an integrating factor\cite{Stone1992,McGlinchey2017} and in the radiation energy density equation the flux divergence and absorption/emission terms are operator split. In 1D Cartesian geometry, the method is as follows:
	\begingroup
	\allowdisplaybreaks
	\begin{align}
	F_{\nu,i+1/2}^{n+1}  &= \exp \left[- \tau_{i+1/2} \right] F_{\nu,i+1/2}^{n} \\
	&-  \frac{c\left(1-\exp \left[- \tau_{i+1/2} \right]\right)}{3dx\hat{\kappa}_{F,\nu,i+1/2}}\left(E_{\nu,i+1}^n-E_{\nu,i}^n\right) \ , \nonumber \\
	E_{\nu,i}^* &= E_{\nu,i}^{n} - dt\frac{F_{\nu,i+1/2}^{n+1}-F_{\nu,i-1/2}^{n+1}}{dx} , \\
	\Delta_{R,\nu,i} &= \left( \frac{S_{\nu,i}}{c} -  E_{\nu,i}^*\right)\left(1-\exp\left[-c\kappa_{P,\nu,i} dt\right]\right) \ , \\ 
	E_{\nu,i}^{n+1} &= E_{\nu,i}^* + \Delta_{R,\nu,i} \\
	U_{e,i}^{n+1} &= U_{e,i}^{n} - \sum_{\nu} \Delta_{R,\nu,i} \ , \\
	\tau_{i+1/2} &= 3 \hat{\kappa}_{F,\nu,i+1/2} c' dt \ , \\
	\hat{\kappa}_{F,\nu,i+1/2} &= \frac{2}{\frac{1}{\kappa_{F,\nu,i}}+\frac{1}{\kappa_{F,\nu,i+1}} } \ ,
	\end{align}
	\endgroup
	with time index $n$, spatial index $i$, cell width $dx$ and time step $dt$ - the above is easily generalised to other geometries and multiple spatial dimensions. The exponential integrating factors are approximated using a combination of Taylor series and fast numerical approximations\cite{Perini2018}.  Externally provided opacity data is needed for both $\kappa_{P,\nu}$ and $\kappa_{F,\nu}$. Currently multigroup Planckian and Rosseland opacities are used for $\kappa_{P,\nu}$ and $\kappa_{F,\nu}$ respectively to ensure accuracy in the diffusive limit. Since Chimera is fully explicit, many sub-cycles of the radiation transport are needed per hydrodynamic time step to ensure stability. To maintain computational efficiency, the opacity table indices are cached for quick interpolation in temperature at fixed density. During sub-cycling, the opacities are also only looked up after a user-defined fractional increase in electron temperature on a per-cell basis. A numerical benchmark between multigroup P$_{1/3}$ and implicit Monte Carlo\cite{Fleck1971} is shown in \cref{fig:FleckCummings}. The radiation boundary conditions\cite{Castor2004} are treated as a Marshak boundary with an incident radiation intensity, $I_\nu$:
	\begin{align}
	\hat{n} \cdot F_\nu &= \langle \mu \rangle \left(c E_\nu - \int_{\hat{n} \cdot \Omega < 0} I_\nu(\Omega) d\Omega\right) \\ 
	&- \int_{\hat{n} \cdot \Omega < 0} \left|\hat{n} \cdot \Omega\right| I_\nu(\Omega) d\Omega \ . \nonumber
	\end{align}
	For an isotropic incident flux, this simplifies to:
	\begin{align}
	\hat{n} \cdot F_\nu &= \langle \mu \rangle c E_\nu - \frac{2\langle \mu \rangle + 1}{4} c E_{\nu,\mathrm{inc}} \ ,
	\end{align}
	where $\hat{n}$ is the surface normal of the boundary, $\langle \mu \rangle$ is the average cosine of the exiting flux (generally set to $1/2$) and $E_{\nu,\mathrm{inc}}$ is the energy density of incoming radiation. This boundary condition can be used to include an external isotropic radiation drive to radiation-hydrodynamics problems such as inertial confinement fusion implosions. 
	
\begin{figure}[htp]
		\centering
		\includegraphics*[width=0.9\columnwidth]{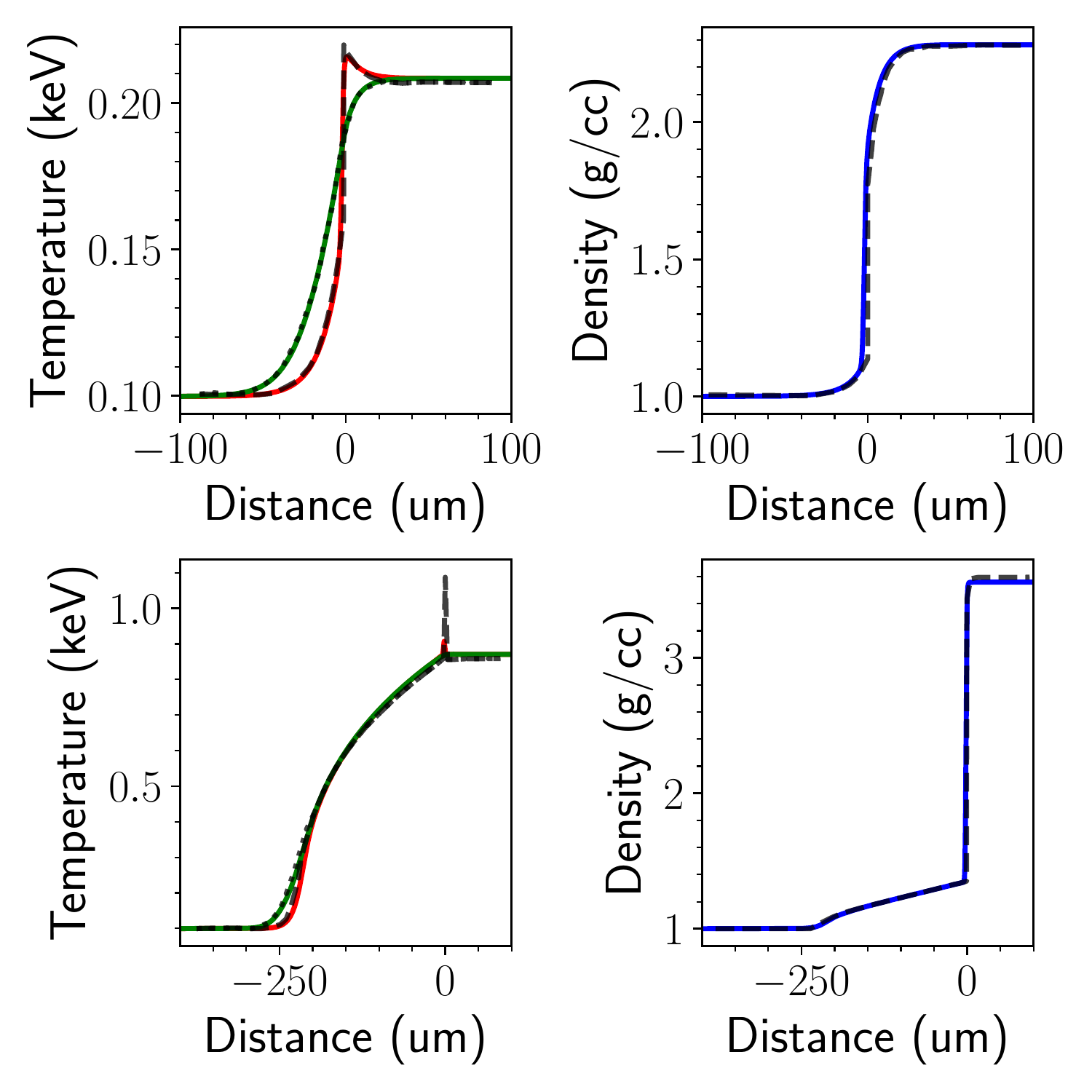}
		\caption{A numerical benchmark of Chimera with the analytic diffusive radiative shock problem of Lowrie \cite{Lowrie2008,Chatzopoulos2019}. The top panels show the Mach 2 subcritical shock and the bottom panels show the Mach 5 supercritical result. The black dashed lines show the analytic results and the red, green and blue lines show the electron temperature, radiation temperature and mass density from Chimera respectively. The plotted Chimera simulations used a cell resolution of $1/3$ um and a reduced speed of light $c' = 3\times10^7$ m/s. }
		\label{fig:LowrieRadShock}
	\end{figure}

	To test the coupling of radiation transport and hydrodynamics, the non-equilibrium radiative shock benchmark of Lowrie\cite{Lowrie2008} is simulated using Chimera. The dimensional upstream and downstream conditions from Chatzopoulos \textit{et al.}\cite{Chatzopoulos2019} are used. The results are shown in \cref{fig:LowrieRadShock}. Good agreement is found between Chimera and the analytic test problem, thus verifying the implementation of the radiation-hydrodynamic coupling in the diffusive regime. Other publications with Chimera simulations\cite{Chittenden2016,Tong2019,McGlinchey2018,Walsh2017,Halliday2022,Crilly2022} have utilised SpK opacity tables and the implementation of radiation transport described here.
	
	\subsection{Cold opacity curves}\label{section:coldopacities}
	
	As discussed in previous sections, many of the approximations used in SpK break down in the high-coupling regime. Therefore, to ensure accuracy in opacity for low temperature, near solid density materials, SpK opacity tables have the option to use Henke\cite{Henke1993} atomic form factors to calculate the photo-absorption opacity:
	\begin{equation}
	\kappa_a = 2 r_0 \lambda f_2 n_i \ ,
	\end{equation}
	where $r_0$ is the classical electron radius, $\lambda$ is the photon wavelength and $f_2$ is the imaginary part of the atomic scattering factor. Around a user-defined temperature the cold opacity is blended into the SpK-calculated opacity:
	\begin{equation}
	\kappa = L(T)\kappa_{\textrm{SpK}}+(1-L(T))\kappa_{\textrm{Henke}} \ ,
	\end{equation}
	where $L(T)$ is the interpolation function and takes a logistic function form.
	
	\subsection{Scattering opacity}
	
	For high photon energies, the scattering cross section can become comparable or exceed the absorption cross section. In the case of isotropic scattering with negligible frequency shift, the radiative transfer equation becomes:
	\begin{align}
	&\left[\frac{1}{c}\frac{\partial}{\partial t} + \Omega \cdot \nabla + (\kappa_\nu+\kappa_{s,\nu})\right]I_\nu = \nonumber \\
	&j_\nu + \kappa_{s,\nu} \cdot \frac{1}{4\pi} \int I_\nu d\Omega \ ,
	\end{align}
	where $\kappa_{s,\nu}$ is the scattering opacity, $I_\nu$ is the spectral radiation intensity and $j_\nu$ is the emissivity. From this, we can form the spectral moment equations including the effects of scattering:
	\begin{align}
	\frac{\partial E_\nu}{\partial t} &= \kappa_{\nu} \left( S_\nu -  c E_\nu\right) - \nabla \cdot \vec{F}_\nu , \\
	\frac{\partial \vec{F}_\nu}{\partial t}  &= - (\kappa_{\nu}+\kappa_{s,\nu}) c \vec{F}_\nu -  c^2 \nabla \cdot \mathbf{P}_\nu \ ,
	\end{align}
	Where $\mathbf{P}_\nu$ is the radiation pressure tensor. From these equations, we can see that the scattering opacity should be included in the flux equation but not the energy density equation. Therefore, the scattering opacity is typically added only to the Rosseland mean or multigroup opacities. In SpK, the total Klein-Nishina cross section is used to calculate the scattering opacity and can optionally be added to the opacity tables. It is worth noting that since we have assumed negligble frequency shift, scattering does not affect the photon-electron energy exchange process. It should be noted that at high photon energies ($h\nu \sim m_e c^2$), Compton scattering is far from isotropic and the frequency shift is large. Therefore, a more detailed approach to scattering is required at these energies.
	
	\section{Post-processing capabilities} \label{section:postprocessing}
	
	Another key application of SpK data is the post-processing of hydrodynamic simulations to produce synthetic diagnostic data. This involves solving the radiation transport equation along a particular line of sight, or characteristic:
	\begin{align}
	I_\nu &= \int_{-\infty}^{\infty} S_\nu \left(1-\exp\left[-\tau_\nu(s) \right]\right) ds \ , \\
	\tau_\nu(s) &= \int^{s}_{-\infty} \kappa_\nu \  ds'  \ ,
	\end{align}
	where $I_\nu$ is the spectral radiation intensity arriving at the detector. These equations can be solved using the same inverse ray trace utilised for neutron transport\cite{Crilly2018}. The spectral grid can be a dynamically evolving treap with SpK running inline or a static array using tables or inline SpK. \Cref{fig:TreapTransport} shows a schematic of a spectrally resolved radiative transfer solution on a Cartesian grid using the treap data structure to dynamically capture the line absorption features:
	\begin{figure}[htp]
		\centering
		\includegraphics*[width=0.99\columnwidth]{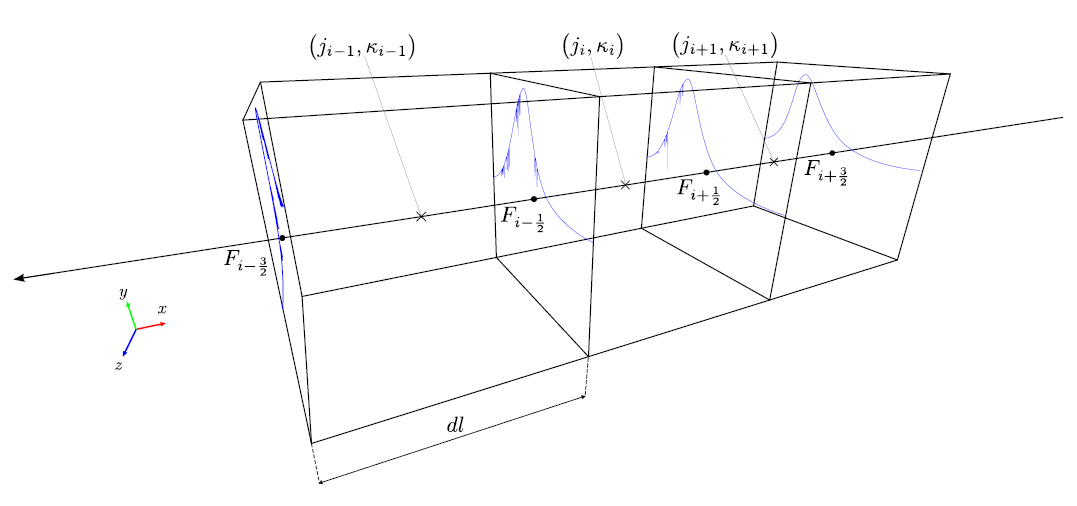}
		\caption{A diagram showing how a spectrum is modified by the cells emissivity, $j_i$, and opacity, $\kappa_i$, as it is marched through the computational grid in the radiation transport solution. The spectrum is stored using the treap data structure, allowing points to be dynamically added as the spectrum evolves. This is especially important when line absorption and emission features emerge from an otherwise smooth spectrum, as demonstrated in this example.}
		\label{fig:TreapTransport}
	\end{figure}
	
	For large 3D simulations $> 10^6$ cells, a separate, more computationally efficient inverse ray trace code based on SpK opacity tables has been written named X-ray Post Processor (XP2).
	
	\section{Discussion \& Future Work}
	
	In this work, we have described the fast atomic and microphysics code SpK. The model works in the chemical picture and solves a modified Saha equation to find bound and free electronic state occupation in chemical equilibrium. These modifications include non-ideal physics such as ionisation potential depression and pressure ionisation and the relaxation of local thermal equilibrium at low electron densities. SpK is then used to compute spectrally resolved opacities on uniquely designed data structures, treaps, and/or compute tabular data for use in radiation-hydrodynamics or post-processing codes. Comparisons with other state-of-the-art codes show reasonable agreement given SpK's range of validity. 
	
	There are many avenues for continued improvement of SpK's capabilities, we give a few notable examples here:
	
	Work is under-way to compute equation of state and transport properties using SpK. These capabilities would operate in much the same way as the opacity model, chemical equilibrium would be solved for and the resultant state and ionic occupancies would be used to compute equilibrium properties. With opacity, equation of state and transport properties all computed from the same model, the radiation-hydrodynamics equations can be closed fully self-consistently.
	
	The use of isolated atom levels and rates allows the potential of reading in more accurate atomic data rather than relying on the screened hydrogenic model with NIST, with the possibility of including fine structure. In this work it has been shown that treating bound states in a plasma as isolated atomic levels with reduced statistical weight and energy shifts can lead to sharp jumps in ionisation at high coupling. As discussed, this is a result of using a low coupling scheme and adding non-ideal physics models to extrapolate to higher coupling. Whether this extrapolation can be improved will be investigated, particularly a consistent coupling between the HNC and Saha models in the dense plasma limit.
	
	The chemical picture allows more complex species to be included in the chemical equilibrium calculation. For example, the formation and dissociation of molecules can be described by analytic partition functions and thus could be included in SpK's extended Saha population solver. The inclusion of molecular states is particularly pertinent in the context of inertial confinement fusion. The deuterium-tritium (DT) fuel is initially in a molecular state and thus accurate simulation of ICF physics requires an adequate description of molecular dissociation.
	
	\section*{Acknowledgements}
	The authors would like to thank Professor Roberto Mancini for many fruitful discussions. This work was supported by the U.S. Defense Threat Reduction Agency (DTRA) under Award No. HDTRA1-20-1-0001.

	\section*{References}
	\bibliography{MuCFRefs}
	
\end{document}